\documentstyle[psfig,epsfig]{article}

\makeatletter

\def\tZ{\tilde{Z}}
\def\tv{\tilde{v}}
\def\tS{\tilde{\Sigma}}
\def\tm{\tilde{m}}

\evensidemargin \oddsidemargin
\makeatother
\newcommand{\be}{\begin{equation}}
\newcommand{\ee}{\end{equation}}
\newcommand{\bea}{\begin{eqnarray}}
\newcommand{\eea}{\end{eqnarray}}
\newcommand{\sgn}{\mbox{sgn}}

\begin{document} 
\noindent
{\footnotesize {\bf To appear in\\[-1.75mm]
{\em Dynamics of Continuous, Discrete and Impulsive Systems}}\\[-1.00mm]
http:monotone.uwaterloo.ca/$\sim$journal}
$~$ \\ [.3in]

\begin{center}
{\large\bf NONLINEAR DYNAMICS IN QUANTUM PHYSICS--THE STUDY OF
QUANTUM CHAOS}

\vskip.20in

L.A.~Caron$^{a}$,
H.~Kr\"{o}ger$^{a}$, X.Q.~Luo$^{b,c}$, \\ G.~Melkonyan$^{a,1} $ 
and K.J.M.~Moriarty$^{d}$  
\\ [2mm]
{\footnotesize  $^{a}$D\'{e}partement de Physique, Universit\'{e} Laval, Qu\'{e}bec, Qu\'{e}bec G1K 7P4, Canada \\ 
$^{b}$CCAST (World Laboratory), P.O. Box 8730, Bejing 100080, China \\ 
$^{c}$Departement of Physics, Zhongshan University, Guangzhou 510275, China \\
$^{d}$Department of Mathematics, Statistics and Computer Science, \\
 Dalhousie University, Halifax, N.S. B3H 3J5, Canada} \\ 

$^1$\footnotesize{Talk given at DCDIS conference in London (Canada) by G. G.
Melkonyan, Email: gmelkony@phy.ulaval.ca}

\end{center}

\noindent Abstract: \\
We use the quantum action to study the dynamics of quantum system at
finite temperature. We construct the quantum action non-perturbatively and find temperature
dependent action parameters. Here we apply the quantum action to study quantum
chaos. We present a numerical study of a classically chaotic 2-D
Hamiltonian system  - harmonic oscillators with anharmonic coupling.  We
compare Poincar\'e sections  for the quantum action at finite
temperature with those of classical action. 


\vskip.2in

\section{\large{\bf Introduction}} 
Chaotic phenomena are found in microscopic physics ruled by quantum mechanics
-- an area presently under very active investigation. For reviews see
Refs.\cite{Blumel, Gutzwiller,Haake,Nakamura,Stockmann}. Examples are the
hydrogen atom in a strong magnetic field \cite{Friedrich} and the quantum
mechanical stadium billiard \cite{McDonald}. Billiard like boundary conditions
have been realized experimentally  in mesoscopic quantum systems, like quantum
dots and quantum corrals, formed by atoms in semi-conductors \cite{Stockmann}. 

Classical chaos theory can not simply be taken over to quantum physics 
(due to Heisenberg's uncertainty relation). Hence workers have tried to
characterize quantum chaos in ways alternative to classical chaos. The
following approaches to quantum chaos  have been considered: 

(i) {\it Gutzwiller's trace formula}. 
Gutzwiller \cite{Gutzwiller} has established a relation between the 
density of states of the quantum system and a sum over 
classical periodic orbits (periodic orbit quantisation). 
The trace formula has been applied in the semi-classical regime (e.g. highly
excited states of atom). Wintgen \cite{Wintgen} applied it to the diamagnetic
hydrogen system and extracted periodic orbit information from experimental
level densities.

(ii) {\it Effective action}. 
Another possibility is to use the conventional effective action
\cite{Jona3,Coleman}. An effective action exists also at finite
temperature \cite{Dolan}. The effective action has been introduced in such a
way that it gives an expectation value $<\phi>=\phi_{class}$ which corresponds
to the classical trajectory and which minimizes the potential energy
(effective potential).  Thus one can obtain the ground state energy of the
quantum system from its effective potential. Because the effective action has
a mathematical structure similar to the classical action, it looks like the
ideal way to bridge the gap from quantum to classical physics and eventually
solve the quantum chaos and quantum instanton problem. Cametti et
al.\cite{Cametti} have  computed in Q.M. the effective action
perturbatively via loop ($\hbar$) expansion and studied quantum chaos in the
2-D anharmonic oscillator.  Perturbation theory forces to terminate such
series at a finite (low) order. The difficulty lies in the estimation of the
remainder.  This poses a problem in the context of chaos, because  chaotic
systems are known to exhibit sensitivity not only to initial conditions but
also to parameters of the system.  There are higher loop corrections to the
effective potential $V^{eff}$ as well as to the mass renormalization $Z$.  The
kinetic term of the effective action is given by an asymptotic series of
higher time derivatives (infinite series of increasing order).  The problem in
interpreting $\Gamma$ as effective action is that those higher time
derivatives require more initial/boundary conditions than the classical
action.

(iii) {\it Bohm's interpretation of quantum mechanics}.
An interesting approach has been taken by Iacomelli and Pettini \cite{Iacomelli}.
They have used Bohm's \cite{Bohm} interpretation of quantum mechanics, which expresses the Schr\"{o}dinger 
equation in polar form. The radial part satisfies a continuity equation of a
velocity field and the phase satisfies a Hamilton-Jacobi equation. The
"trajectories" of the quantum system are equivalent to Lagrangian trajectories
of a fluid velocity field. This allows to apply the concept of classical
chaos theory and in particular to compute Lyapunov exponents. In
Ref.\cite{Iacomelli} this framework has been used to study the chaotic
behavior of the hydrogen atom in an external magnetic field and different
regimes of quantum chaoticity have been observed. 

 (iv) {\it A conjecture by Bohigas et al.} \cite{Bohigas} states that the
signature of a classical chaotic system,  described by random matrix theory,
is a spectral density following a Wigner distribution.

Strictly speaking, there is no universally accepted 
definition of quantum chaos. Many roads have been travelled searching for
"irregular" signals in quantum systems. In the quest for quantum chaos, it has
been popular to examine systems which exhibit classical chaos. However,
Jona-Lasinio et al. \cite{Jona1}  found chaotic behavior in a quantum system
without classical counterpart, i.e., a quantum many-body system undergoing
multiple resonant tunneling. Jona-Lasinio and Presilla \cite{Jona2} further
studied  quantum many-body systems of interacting bosons in the thermodynamic
limit.  They found chaotic signals when the interaction strength drives the
system away from integrability.


However, the study of quantum chaos lacks from the following shortcomings:
(i) In quantum chaos, we have no "local" information of the degree of chaoticity, 
being available in classical chaos via {\it Lyapunov exponents} and Poincar\'e
sections from phase space. (ii) Also little is known about the role of
{\it temperature} in quantum chaos. For example, the analysis of level densities is
insensitive to temperature. To overcome those problems,
we adopt the strategy of building a bridge between quantum physics and classical physics. 
This means a relation involving the quantum transition amplitude and the
classical action. 
 
In the section ~\ref{QA} we present the definition of the quantum action
\cite{Jirari01a}. In the section ~\ref{methods} we describe methods of
construction of the quantum action. The dynamics in 2-D anharmonic oscilator by
use of the quantum action is considered in the section~\ref{chaos}.

\section{\large{\bf Quantum action \label{QA}}}

\begin{figure}[btp]
\begin{center}\leavevmode
\includegraphics[width=0.6\linewidth, height=0.3\linewidth]{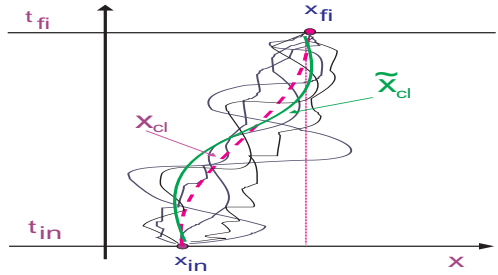}
\caption{Path integral takes into account all possible paths between two
boundary points $x_{in}$, $x_{fi}$ }  \label{TransAmpl}\end{center} 
\end{figure}

\noindent Feynman's path integral formulation of quantum mechanics
~\cite{Feynman,Schulman} is helpful in introducing the ideas of the quantum
action. In this formulation the quantum mechanical transition amplitude or
Greens function is defined as  \be
\label{Feyn}
G(x_{fi},t_{fi},x_{in},t_{in})=\int d[x]\mbox{ } exp{\left(\frac{i}{\hbar} S[x]\right)},
\ee   
where $S$ is classical action, $ x_{in}$, $x_{fi}$ are initial and
final boundary points, $t_{in},t_{fi}$ are initial and final times. The
integral is understood as an integral over all possibles paths connecting
initial and final boundary points (see Figure ~\ref{TransAmpl}).
   In some exceptional cases
(harmonic oscillator and free particle) integral~(\ref{Feyn}) can be presented as
 \be
\label{harmonic}
G(x_{fi},t_{fi},x_{in},t_{in})=Z\mbox{ }exp{\left. \left( \frac{i}{\hbar} 
S[x_{cl}]\right) \right|_{x_{in},t_{in}}^{x_{fi},t_{fi}}  }, \ee   
where $x_{cl}$ is the classical path which minimizes the classical action and
solves Euler--Lagrange equations of motion.  The definition of the
quantum action   \cite{Jirari01a} is guided by Equation ~(\ref{harmonic}).
\\  \noindent {\it Conjecture}: 
For a given
classical action $S = \int dt \frac{m}{2} \dot{x}^{2} - V(x)$  there is a
quantum action  $\tilde{S} = \int dt \frac{\tilde{m}}{2} \dot{x}^{2} -
\tilde{V}(x)$,  which allows to express the quantum mechanical transition amplitude $G$ as %
\begin{equation} \label{DefQuantumAction}
G(x_{fi},t_{fi}; x_{in},t_{in}) = \tilde{Z} 
\exp \left. \left(  \frac{i}{\hbar} \tilde{S}[\tilde{x}_{cl}]\right) 
\right|_{x_{in},t_{in}}^{x_{fi},t_{fi}}  .
\end{equation}
Here $\tilde{x}_ {cl}$ denotes the classical path, such that the
action $\tilde{S}[\tilde{x}_{cl}]$  is minimal (we exclude the occurrence of
conjugate points or caustics).  $\tilde{Z}$ denotes the normalization factor
corresponding to $\tilde{S}$.  Eq.(\ref{DefQuantumAction}) is valid with 
the {\em same} action $\tilde{S}$ for all sets of 
boundary points $x_{fi}$, $x_{in}$ for a given time interval $T=t_{fi}-t_{in}$. 
The parameters of the quantum action depend on the time $T$.  
Any dependence on $x_{fi}, x_{in}$ enters via the trajectory 
$\tilde{x}_ {cl}$. $\tilde{Z}$ depends on the action parameters and $T$, 
but not on $x_{fi}, x_{in}$. 

 This action has a mathematical
structure like the classical action, and takes into account quantum effects
(quantum fluctuations) via renormalized action parameters. The quantum action
has the following attractive properties \cite{Jirari01b}: (i) It can be
computed non-perturbatively, via numerical simulations.  (ii) It is defined
for arbitrary finite temperature. (iii) In the zero-temperature limit 
analytic expressions exist for the quantum action, provided that the classical
potential is sufficiently symmetric (parity symmetric in D=1 and rotationally
symmetric in D=2,3).  (iv) In a numerical study of the 1-D anharmonic
oscillator,  we searched for the presence of higher order time derivative
terms in the kinetic term of the quantum action. No signal for such term was
found \cite{Jirari01b}, in contrast to the effective action formalism, where
higher derivative terms persist.  The quantum action has proven useful for a
rigorous definition and  quantitative computation of quantum instantons~\cite{Jirari01c}.

The quantum action at finite time $T$ can be interpreted as action at 
finite temperature.
The partition function in quantum mechanics requires to go over from real time to imaginary time ($t=-it$) 
and to impose periodic boundary conditions \cite{Kapusta}.
In Ref.\cite{Jirari01b} we have shown that the expectation value of a quantum mechanical observable $O$ at thermodymical 
equilibrium can be expressed in terms of the quantum action along its
classical trajectory from $x_{in}$, $\beta_{in}=0$ to $x$, $\beta$, given by %
\bea
&& \tilde{\Sigma}_{\beta} 
\equiv 
\tilde{S}_{\beta}[\tilde{x}_{cl}]|_{x_{in},0}^{x,\beta} = 
\int_{0}^{\beta} d \beta' ~ \frac{m}{2 \hbar^{2}} (\frac{d \tilde{x}_{cl}}{d \beta'})^{2} 
+ \tilde{V}(\tilde{x}_{cl})
\nonumber \\
&& \frac{\delta \tilde{S}_{\beta}}{\delta x}[\tilde{x}_{cl}] = 0 
\nonumber \\
&& \tilde{x}_{cl}(\beta_{in})=x_{in}, ~ \tilde{x}_{cl}(\beta) = x ~ ,
\eea
where $T$ denotes the absolute value of imaginary time. The parameters $\beta$, temperature $\tau$ and $T$ are related by 
$\beta = \frac{1}{k_{B} {\tau}} = T/\hbar$.
Numerical studies with the quartic potential in 1-D have shown that the
parameters  of the quantum action are temperature dependent
\cite{Jirari01a,Jirari01c,Jirari01b}. 

\section{Methods of construction of the quantum action \label{methods}}
\subsection{Analytic form of quantum action at zero temperature}
\label{sec:AnalyticForm}

From the experimental point of view, the ground state properties of a quantum
system are often much better  known than excited states. Zero temperature
corresponds to large time and  from the point of view of chaos theory, 
it is interesting to study dynamical behavior at large times. 
Examples:
(a) Particles moving in storage rings. 
(b) In classical Hamiltonian chaos, there are quasi-periodic trajectories,
remaining for a long time in the vicinity of regular islands. What is the
corresponding behavior in quantum mechanics? 
(c) What is the quantum analogue
of the KAM theorem? 
(d) The construction of Poincar\'e sections and Lyapunov
exponents involves  long time trajectories. Likewise the quantum analogue of
Poincar\'e sections and Lyapunov  exponents will require long time evolutions.

In the following we work in imaginary time. 
In order to avoid notational confusion, we continue to use the time parameter
$T$. In particular, $T \to 0$ denotes the high temperature limit, where the
quantum action approaches its classical counter part. $T \to \infty$ denotes
the zero-temperature limit, where the quantum action is determined by the
Feynman-Kac formula.   In Ref.\cite{Jirari01b} the following analytic
transformation law at temperature zero has been derived in 1-D relating the
classical action $S=\int dt \frac{1}{2} m \dot{x}^{2} + V(x)$ to the quantum
action $\tilde{S} = \int dt \frac{1}{2} \tilde{m} \dot{x}^{2} + \tilde{V}(x)$,
 \be \label{TransformFeynKac} 2 m(V(x) - E_{gr}) =   2
\tilde{m}(\tilde{V}(x)-\tilde{v_0})-\frac{\hbar}{2}\frac{\frac{d}{dx} 2
\tilde{m}(\tilde{V}(x)-\tilde{v_0}) }{\sqrt{2
\tilde{m}(\tilde{V}(x)-\tilde{v_0})}}\sgn (x).
\ee
This differential equation relates the classical action to the quantum action
at zero temperature limit.
Moreover, the following relation between the
ground state wave function and the quantum action holds %
\be
\label{GroundStateFeynKac}
\psi_{gr}(x) = \frac{1}{N} ~ e^{ - \int_{0}^{|x|} dx' ~ 
\sqrt{2 \tilde{m}( \tilde{V}(x') - \tilde{v}_{0} ) }/\hbar }, ~
E_{gr} = \tilde{v}_{0} ,
\ee
Eqs.(\ref{TransformFeynKac},\ref{GroundStateFeynKac}) are valid under the assumption 
 of a parity symmetric potential $\tilde{V}(x)$ with a unique minimum at
$x=0$, $\tilde{V}(0)=\tilde{v}_{0}$. Similar laws hold for rotationally
symmetric potentials in D=2 and 3. For a number of systems, where the ground
state energy and wave function is known analytically, it has been shown
\cite{Jirari01b} that the quantum action reproduces exactly the ground state
energy and wave function.  Examples in 1-D are: Monomial potentials and the
inverse square potential.  Examples for radial motion in 3-D are: Monomial
potentials and Coulomb potential (hydrogen atom) in the sector of non-zero
angular momentum. In particular, the ground state energy coincides with the
minimum of the quantum potential $\tilde{V}$ and the location of the minimum
of the quantum potential  coincides with the location of the maximum of the
wave function. One should note that the transformation law,
Eq.(\ref{TransformFeynKac}), determines mass $\times$ potential, but not both,
mass and potential individually. The determination of the renormalized mass
requires a non-perturbative numerical calculation. 

\subsection{\large{\bf Calculation of quantum action by fit to the quantum
mechanical transition amplitude}}  

A method to obtain the quantum action follows directly from the definition of 
the quantum action. Equation (\ref{DefQuantumAction})
must hold for all boundary points for a given time interval $T=t_{fi}-t_{in}$.
A parametrization of the quantum action as a polynomial by taking into account
symmetry of the system is very useful. It permits to fit the quantum action to
the quantum mechanical transition amplitude. After parametrization one
minimizes the sum of square differences  \be \label{LeastSquares}
\chi^2=\sum_{x_{in},x_{fi}}\left|G(x_{fi},T,x_{in},0
)-\tilde{Z}\mbox{ } exp\left[ \frac{i}{\hbar} \tilde{S} \right] \right|^2, \ee 
to calculate the parameters of the quantum action. $G(x_{fi},T,x_{in},0)$ is the exact quantum mechanical
transition amplitude obtained from the solution of the Schr\"odinger equation. 

\subsection{\large{\bf Calculation of quantum action by renormalization group
flux equation}}

In quantum field theory action parameters depend on cut--off parameter
$\Lambda$ or on spatial and time lattice spacing. This dependence is governed by a
renormalization group equation. In constructing the quantum action we are
close to the continuum limit. As scale and temperature dependence of the
action are similar, one can write an equation which governs the action
parameters when temperature is varied ~\cite{Jirari01b}.
 Following our conjecture
this equation reads
\be
\label{FluxEq}
-\frac{1}{\tZ_\beta}\frac{d
\tZ_\beta}{d\beta}+\frac{\tS_\beta}{d\beta}+\frac{\hbar^2}{2m}
\left[\left( \frac{d\tS_\beta}{d x}\right)^2-
\frac{d^2\tS_\beta}{d^2x}\right]-V = 0.   \ee 
$\tS$ is given by the quantum action along its classical  trajectory
and $\beta=T/\hbar$ as before.
\be
\tS_\beta=\tilde{S}_\beta[\tilde{x}_{cl}] =\left.\int_0^{\beta} dt
\frac{ \tilde{m} } {2\hbar^2} \dot{ \tilde{x}}_{cl}^2
+\tilde{V} ( \tilde{x}_{cl} )\right|_{x_{in}}^{x_{fi}}.
\ee

By parametrizing the quantum potential in terms of $\tv_{k}|x|^{k} $ 
corresponding to the symmetry of $\tilde{V}(x)$ from equation~(\ref{FluxEq}) we have a system of
linear differential equations for the quantum action parameters
$\tv_k, \tm $ and $\tZ $.
\be
\tS_\beta=\tS_\beta\left[ \tm(\beta), \tv_0(\beta),\tv_1(\beta),\dots, x,
\beta\right]. \ee
One can solve this system of differential equations either by giving initial
conditions at $\beta=0$ (classical limit) or at $\beta=\infty$. Like
equation~(\ref{TransformFeynKac}) this system of differential equations
connects the classical and quantum potentials. It determines the flow of the
action parameters as temperature is lowered or increased. 

\subsection{Numerical results} 
\begin{figure}[tp]
\begin{center}\leavevmode
\includegraphics[width=0.6\linewidth,angle=-90]{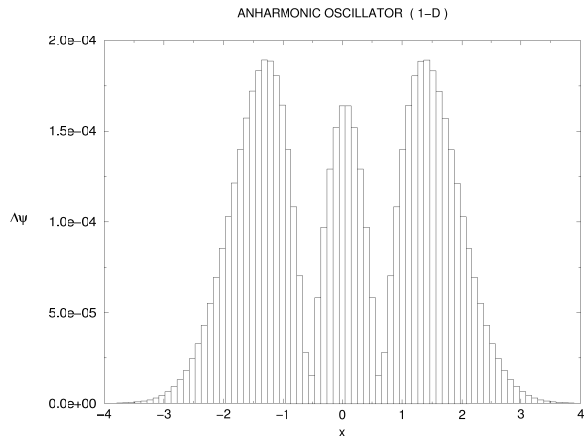}
\caption{ 1-D anharmonic oscillator. Differences between ground state wave
functions obtained from Schr\"odinger equation and from quantum action. } 
 \label{HarmonicWF}\end{center}  \end{figure}

First we will try to estimate the precision of numerical calculations  of
the quantum action in the zero-temperature limit (long time limit) by
contrasting it with the analytic expressions of 
Eqs.(\ref{TransformFeynKac},\ref{GroundStateFeynKac}). We consider in 1-D the
classical action %
\bea
S &= \int dt \frac{m}{2} \dot{x}^{2} + V(x), ~~~ 
V(x) = v_{2} x^{2} + v_{4} x^{4},& \nonumber \\ 
 &m = 1,
\mbox{ } v_{2} = 1,  \mbox{ }
v_{4} = 0.01 ~ .&
\eea
The Q.M. Hamiltonian has the following ground state energy, 
\be
\label{ExactEnergy}
E_{gr}= 0.710811 ~ .
\ee
For the quantum action we make an ansatz
\be
\tilde{S} = \int dt \frac{\tilde{m}}{2} \dot{x}^{2} + \tilde{V}(x), ~~
\tilde{V}(x) = \tilde{v}_{0} + \tilde{v}_{2} x^{2} + \tilde{v}_{4} x^{4} + \tilde{v}_{6} x^{6} ~ .
\ee
From a fit of the quantum action to Q.M. transition amplitudes 
(expression~(\ref{LeastSquares}), for more details see Ref.\cite{Jirari01a}) we find the action parameters at $T=4.5$
\bea
\label{1DNumData}
& \tilde{m} = 0.9990(7), \mbox{ }
 \tilde{v}_{0} = 0.79863(1), \mbox{ }
 \tilde{v}_{2} = 1.013(2),& \nonumber  \\
& \tilde{v}_{4} = 0.0099(9), \mbox{ }
\tilde{v}_{6} = 0.0000(3) ~ .& \nonumber
\eea
The parameters $\tilde{m}$, $\tilde{v}_{2}$, $\tilde{v}_{4}$, $\tilde{v}_{6}$ have asymptotically stabilized at about $T=4.5$. 
However, the parameter $\tilde{v}_{0}$ has a slow fall-off behavior for $T \to \infty$. For large $T$ we have fitted $\tilde{v}_{0}$ to the numerical data by                                                          %
\be
\tilde{v}_{0}(T) \leadsto_{T \to \infty} A + \frac{B}{T} + \frac{C}{T^{2}} ~ ,
\ee
and find $A=0.710819$. Thus the asymptotic extrapolation $\tilde{v}_{0}(T \to \infty) = 0.710819$ 
is very close to $E_{gr}$, given by Eq.(\ref{ExactEnergy}). 
This confirms $E_{gr}=\tilde{v}_{0}$, predicted by Eq.(\ref{GroundStateFeynKac}), 
i.e. the ground state energy is given by the minimum of the quantum potential.
Let us further check the consistency of other parameters of the quantum potential. This can be done by using 
the transformation law Eq.(\ref{TransformFeynKac}).
Because this transformation law does not determine 
$\tilde{m}$, we have taken $\tilde{m}$ from the numerial data.
The transformation law then determines the parameters of the quantum potential.
We find,
\bea
\tilde{v}_{2} &=& \frac{2}{\tilde{m}} [ m E_{gr}/\hbar]^{2} = 1.01151 
\nonumber \\
\tilde{v}_{4} &=& \frac{2}{3} \frac{\sqrt{2 \tilde{m} \tilde{v}_{2}}}{\hbar} ~ 
[ \tilde{v}_{2} - v_{2} \frac{m}{\tilde{m}} ] = 0.009967 
\nonumber \\
\tilde{v}_{6} &=& \frac{1}{4} \frac{\tilde{v}_{4}^{2}}{\tilde{v}_{2}}
+ \frac{2}{5} \frac{\sqrt{2 \tilde{m} \tilde{v}_{2}}}{\hbar} ~ 
[ \tilde{v}_{4} - v_{4} \frac{m}{\tilde{m}} ]
= 2.89 ~ 10^{-7} ~ .
\eea
This is consistent with the numerical results, given by Eq.(\ref{1DNumData}). 
Finally, let us look at the ground state wave function. 
As reference solution, we have solved the Schr\"{o}dinger equation 
with the original Hamiltonian. Secondly, we have computed the wave function 
from the quantum action via Eq.(\ref{GroundStateFeynKac}), using the numerical data given by Eq.(\ref{1DNumData}).
The comparison is given in Figure~\ref{HarmonicWF}, showing good agreement.

\begin{figure}[tp]
\begin{center}\leavevmode
\includegraphics[width=0.9\linewidth, height=0.45\linewidth]{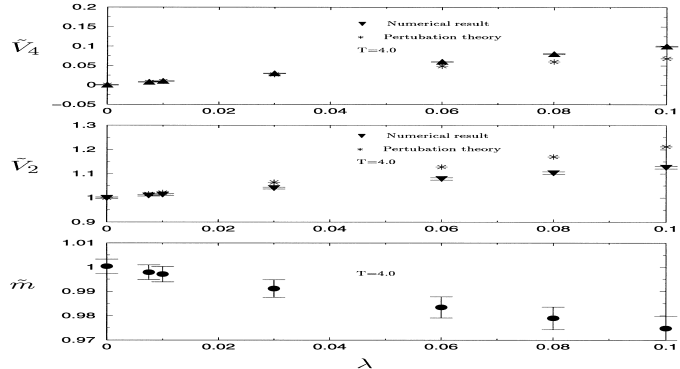}
\caption{Differences between quantum action parameters and effective action.
Potential $V=0.5x^2+\lambda x^4$.}  
\label{QuantumVersusEffective}\end{center}   \end{figure}

\begin{figure}[tp] 
\begin{center}\leavevmode \includegraphics[width=0.9\linewidth,
height=0.55\linewidth]{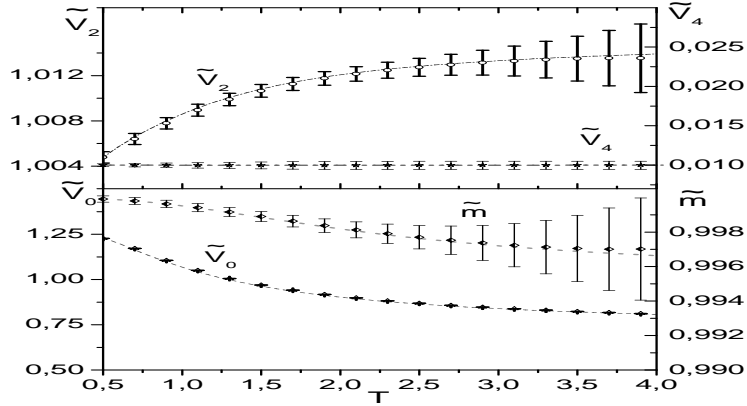} \caption{Quantum action parameters obtained
from the direct fit and from solution of the flux equation. Potential
$V=0.5x^2+0.01x^4$. Lines: flux equation and symbols: fit. 
}   \label{DirectFlux}\end{center}  \end{figure}

Now let us consider the quantum action at finite temperature. The results obtained 
by minimizing the error in the expression~(\ref{LeastSquares})
(direct fit) show that at high temperature parameters of quantum and
classical actions coincide and at low temperature the parameters of the
quantum action are saturated~\cite{Jirari01a}. Figure~\ref{QuantumVersusEffective}
presents a comparison of the quantum action with the conventional effective
action.
We see at small coupling that the quantum and effective action are
similar but at high values of coupling there is some discrepancy.
Finally, we presnt a comparison of the methods to determine the quantum
action-- via fit (\ref{LeastSquares}) and via solution of the renormalisation
group flux equation~(\ref{FluxEq}). Such comparison is presented in
Figure~\ref{DirectFlux}. Two different methods reproduce the same
result. 

In summary of this section, we have shown for the zero-temperature limit
(regime of the Feynman-Kac formula) 
that the transition amplitude can be parametrized by the quantum action.
A comparison of analytical versus numerical computation of the quantum action has shown good agreement.
We have shown that in effective and quantum actions can coincide at small
coupling constants. Also we show that solutions obtained by two different
methods (direct fit to quantum mechanical transition amplitude and flux
equation) agree very well.

\section{\large{\bf Quantum chaos in 2-D anharmonic oscillator \label{chaos}}}
\noindent As is well known 1-D conservative systems with a time-independent Hamiltonian are integrable and do not display classical chaos.
An interesting candidate system to display chaos is the K-system,
corresponding to the potential $V=x^{2}y^{2}$. It describes a 2-D Hamiltonian
system, being almost globally chaotic, having small islands of stability
\cite{KSystem}. However, according to a suggestion by Pullen and Edmonds
\cite{Pullen}, it is more convenient from the numerical point of view to
consider the following system, which also exhibits classical chaos. It is a
2-D anharmonic oscillator, defined  by the following classical action,  %
\bea
&S = \int_{0}^{T} dt ~ \frac{1}{2} m (\dot{x}^{2} + \dot{y}^{2})  + V(x,y),
~ V(x,y) = v_{2}(x^{2} + y^{2}) + v_{22} x^{2}y^{2} & \nonumber \\ 
&m = 1,\mbox{ } v_{2} = 0.5, \mbox{ } v_{22} =0.05 ~ .& 
\eea
We work in imaginary time and the convention $\hbar=k_{B}=1$ is used.
For the corresponding quantum action, we make the following ansatz,
compatible with time-reversal symmetry, parity conservation and symmetry under exchange $x \leftrightarrow y$,
\begin{eqnarray}
\tilde{S} &=& \int_{0}^{T} dt ~ 
\frac{1}{2} \tilde{m} (\dot{x}^{2} + \dot{y}^{2}) 
+ \tilde{V}(x,y), ~~~
\nonumber \\
\tilde{V} &=& \tilde{v}_{0} 
+ \tilde{v}_{2} (x^{2} + y^{2}) 
+ \tilde{v}_{22} x^{2}y^{2} 
+ \tilde{v}_{4} (x^{4} + y^{4}) ~ .
\end{eqnarray}
We have determined numerically the parameters of the quantum action 
for transition times from $T=0$ up to $T=4$, where a regime of asymptotic stability is reached. For $T=4$ corresponding to temperature $\tau=0.25$, 
we found 
\begin{eqnarray}
&\tilde{m} = 0.99814(6),\mbox{ }
\tilde{v}_{0} = 1.29373(2),\mbox{ } 
\tilde{v}_{2} = 0.5154(3)&
\nonumber \\ 
&\tilde{v}_{22} = 0.0497(1),\mbox{ } 
\tilde{v}_{4} = -.0008(7) ~ .&
\end{eqnarray} 
We have also included terms $\dot{x} \dot{y}$, $xy$, $xy^{3}+x^{3}y$.
Their coefficients were found to be very small (compared to machine precision). 
Also we included terms $x^{2}y^{4} + x^{4}y^{2}$, $x^{4}y^{4}$. Those coefficients were found compatible with zero within error bars.
The quantum action slightly modifies the parameters $\tilde{v}_{2}$ and the parameter $\tilde{v}_{22}$. 

Now let us look at the chaotic behavior of the classical and the quantum 
action. We have computed Poincar\'e sections from the classical action 
and from the quantum action for a variety of temperatures. 
The equations of motion of the quantum action have been solved using a 4-th order Runge-Kutta algorithm, and Henon's algoritm was used to compute the Poincar\'e sections. 
Let us discuss what we expect.
First, the classical system under consideration is 
chaotic due to the anharmonic 
interaction term. This term is small compared to the harmonic 
oscillator term. This property proliferates to the quantum potential.
Here "smallness" is meant in the following sense: The quantum system at 
zero temperature is determined by the ground state properties. 
We have computed 
$r_{0}=\langle r \rangle_{gr}$, the radius of the ground state wave 
function (corresponding to the Bohr radius $a_{0}$ in the hydrogen atom).
We obtain $r_{0} = 0.8766$. 
It turns out that the anharmonic term is smaller than the harmonic term 
for any length $r < r_{0}$.
Thus one expects roughly the same "amount" of chaotic behavior in the 
quantum system as in the classical system. 
Second, renormalization (i.e. change of action parameters when going from 
the classical action to the quantum action) tends to increase harmonic terms, 
i.e. drive the system to a Gaussian fixed point. At a Gaussian fixed point the system is integrable. i.e. non-chaotic.
This effect has been observed in a number of numerical studies in 1-D systems \cite{LACaron,Jirari01a,Jirari01c}. In a system with a symmetric double well 
potential this effect was found to lead to a softening of the double well shape of the quantum potential (lower barrier, more narrow minima, 
even vanishing of the double well). 
This translates to a softening or even an "evaporation" of quantum instantons 
compared to the classical instantons \cite{Jirari01c}. 
This effect is present also in the system under consideration here. We find
$\delta \tilde{v}_{2} = \tilde{v}_{2} - v_{2} = 0.0154(3)$ compared to 
$\delta \tilde{v}_{22} = \tilde{v}_{22} - v_{22} = -.0003(1)$. 
We expect that this effect may lead to a slight softening of the chaotic 
behavior in the quantum system.
Third, a weak anharmonic term means that the quantum potential reaches 
a regime of asymptotic stability as function of $T$ already for small $T$ 
($T \approx 4$). That is, when going from high to low temperature, 
already at relatively high temperature the asymptotic regime (Feynman-Kac limit) is reached. 
\begin{figure}[tp]
\begin{center}
\leavevmode
\includegraphics[width=0.48\linewidth, height=0.6\linewidth]{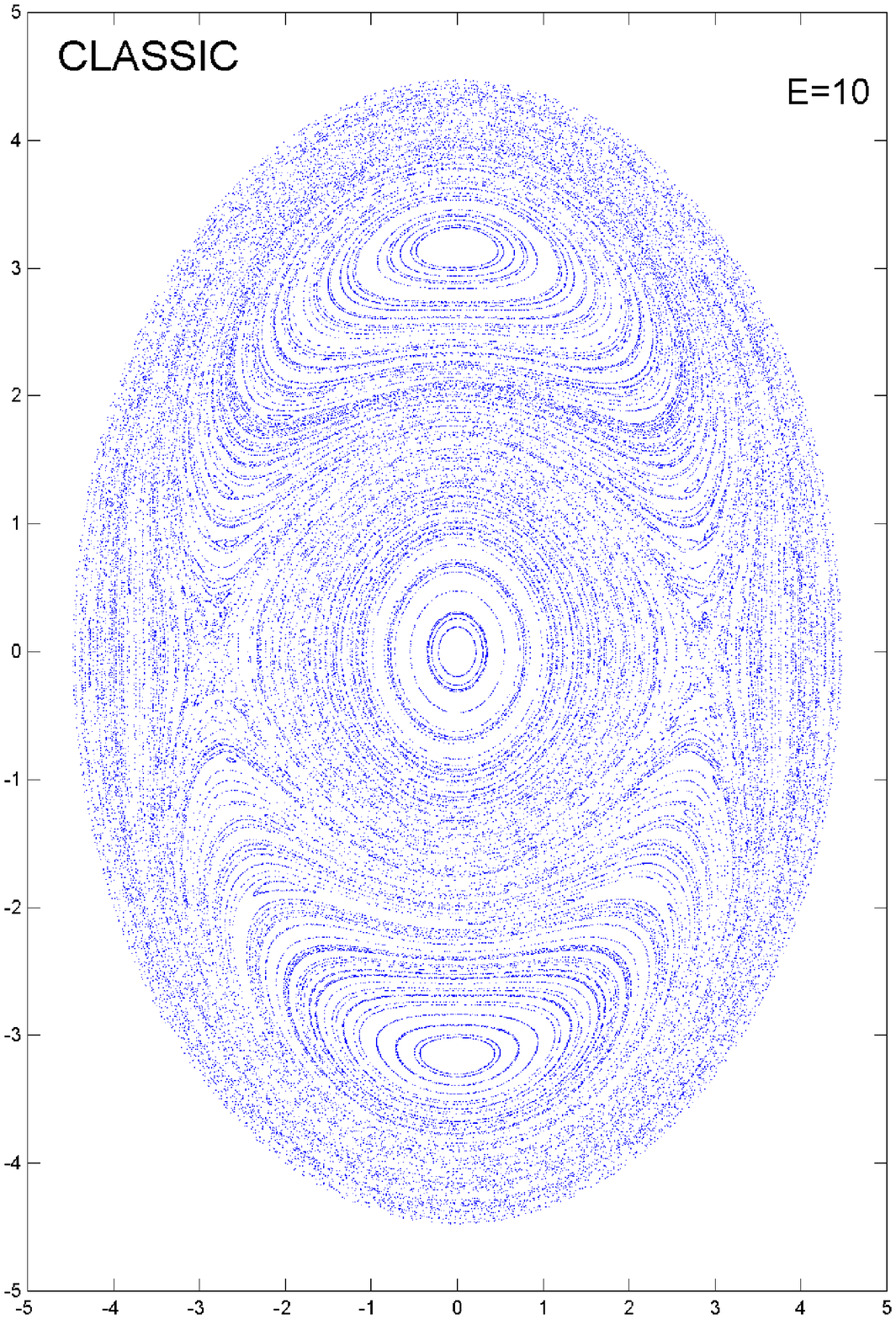}
\includegraphics[width=0.48\linewidth, height=0.6\linewidth]{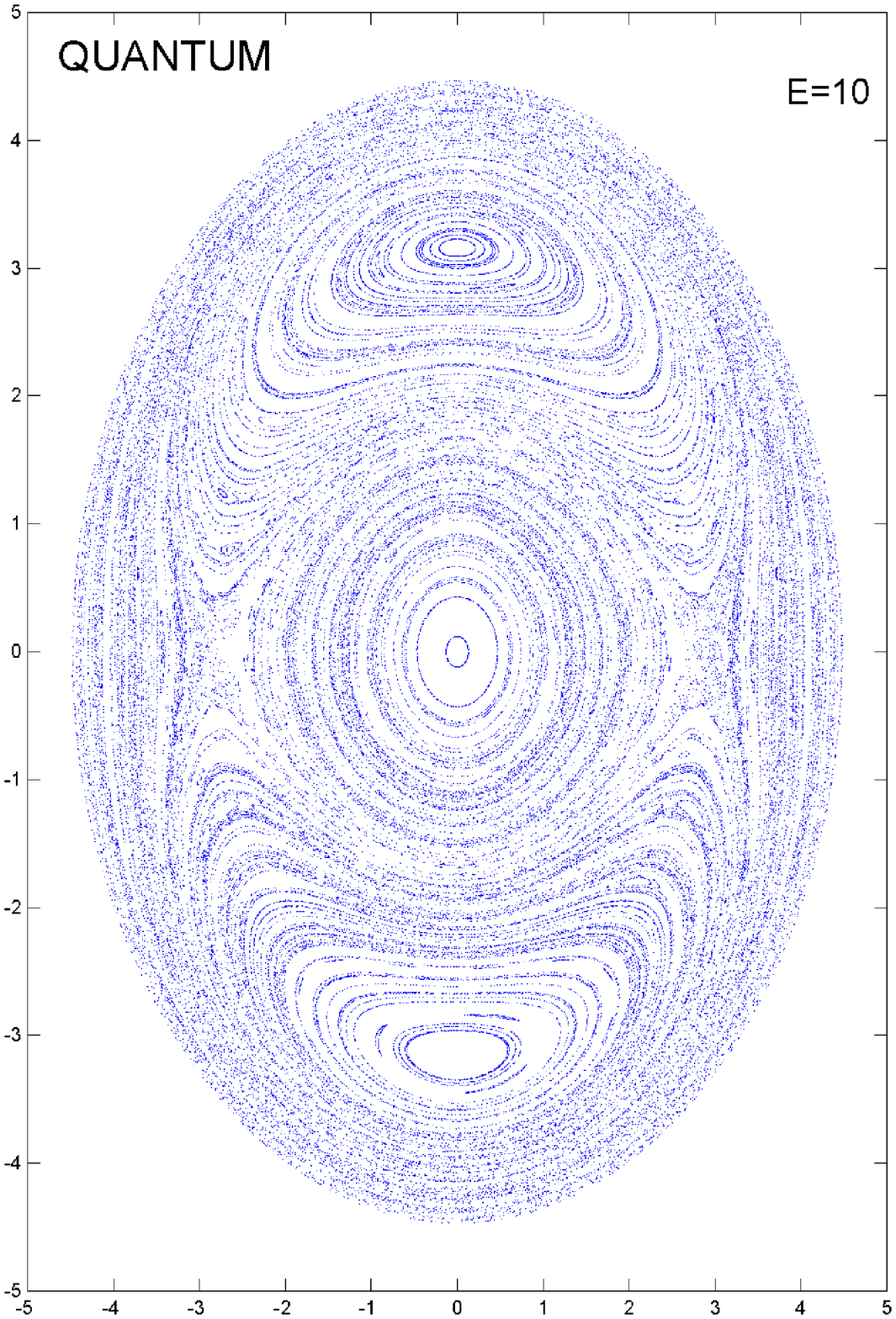}
\caption{2-D anharmonic oscillator. Poincar\'e sections at energy $E=10$. 
Left part classical action, right part quantum action at temperature $\tau=0.25$. }  
\label{E10}
\end{center}  
\end{figure}

\begin{figure}[tp]
\begin{center}
\leavevmode
\includegraphics[width=0.48\linewidth, height=0.6\linewidth]{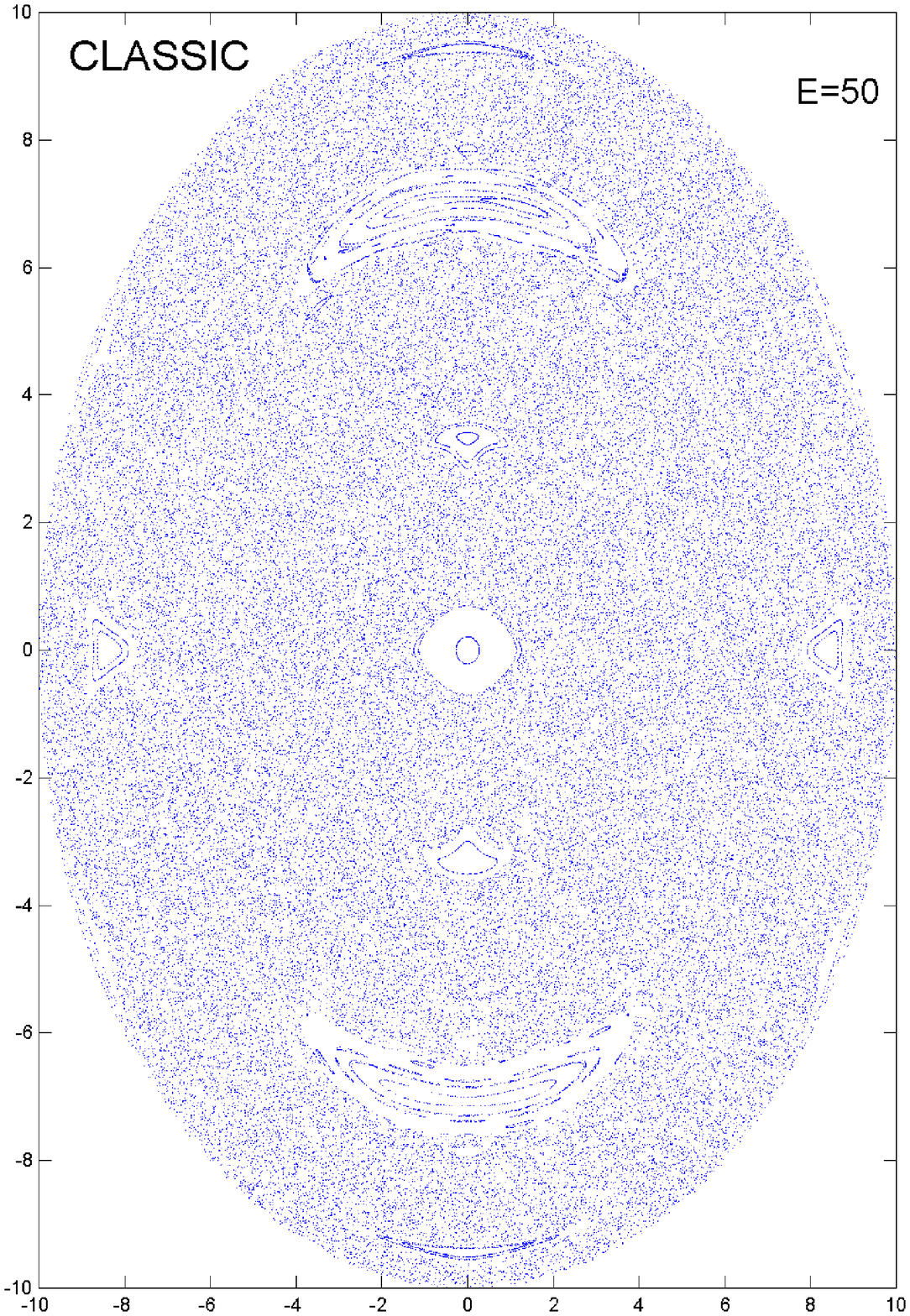}
\includegraphics[width=0.48\linewidth, height=0.6\linewidth]{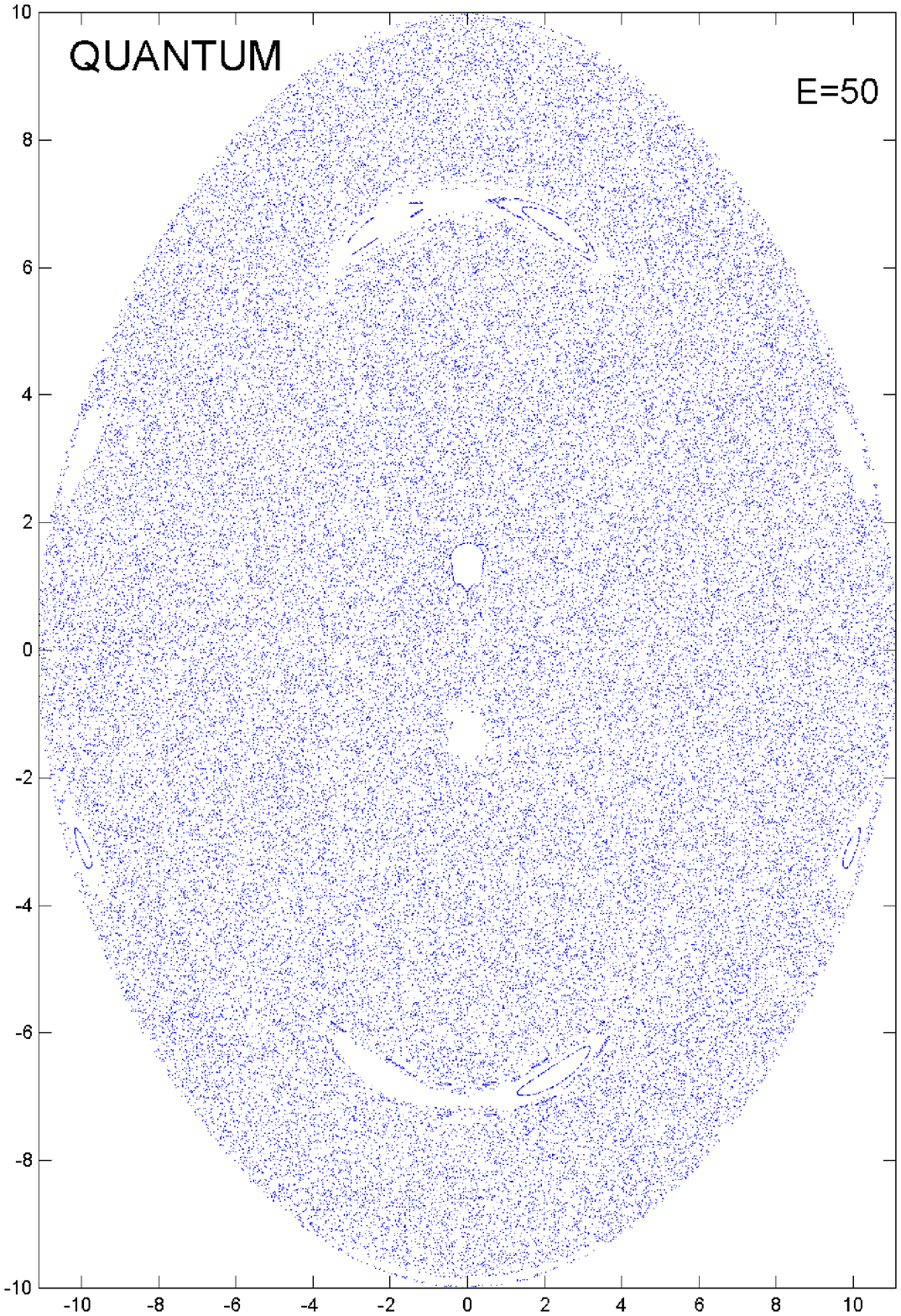}
\caption{ Same as Figure~\ref{E10}, but $E=50$}  
\label{E50}
\end{center}  
\end{figure}

We computed the Poincar\'{e} sections for the quantum action at temperature $\tau=0.25$ (corresponding to $T=4$) 
for energies $E=10, 20, 50$. 
One should keep in mind that the classical action at $T=0$ is equivalent to a quantum action at temperature $\tau=\infty$.
Poincar\'e sections compared to their counter part from the classical action for energies $E=10, 50$ are shown in 
Figures~\ref{E10} and~\ref{E50}. One observes that the quantum system also
displays chaos, and the Poincar\'{e} sections are slightly different from
those of the classical action. Like in the classical case also in the quantum
system the "amount" of chaoticity increases with increasing energy.   
Moreover, the difference between classical and quantum Poincar\'e sections
becomes more accentuated with increase of energy.

\section{Summary}
\noindent We have discussed the use of the quantum action, which can be 
considered as a renormalized classical action at finite temperature and methods of its construction. Based on the quantum action we suggest a proper definition 
of quantum chaos and explore it numerically.
As example, we have considered in 2-D the harmonic oscillator with a weak anharmonic coupling and computed the quantum action at 
various temperatures. We compared Poincar\'{e} sections 
of the quantum action with those of the classical action and found some difference in their chaotic behavior. \\

\noindent {\bf Acknowledgements} \\ 
H.K. and K.D.M.M. are grateful for support by NSERC Canada. 
X.Q.L. has been supported by NSF for Distinguished Young Scientists of China, by Guangdong Provincial NSF and by the Ministry of Education of China. G.M. has been supported by the Quebec Ministry of Education.

\end{document}